\documentclass[aps,pra,twocolumn,showpacs]{revtex4-1}
\usepackage[colorlinks,bookmarks=false,citecolor=blue,linkcolor=red,urlcolor=black]{hyperref}
\usepackage{graphicx}
\usepackage{amssymb,amsmath} 
\usepackage{amsfonts}
\usepackage{color}
\usepackage[utf8]{inputenc}
\usepackage[inline]{enumitem}
\usepackage[T1]{fontenc}
\usepackage{ulem}
\usepackage[dvipsnames]{xcolor}

\usepackage{latexsym}

 \def\urlprefix{}
 \def\url#1{}
\def\f{\mathrm{f}}
\def\frozen{\emph{frozen}}
\def\be{\begin{equation}}
\def\ee{\end{equation}}
\def\bea{\begin{eqnarray}}
\def\eea{\end{eqnarray}}
\def\bi{\begin{itemize}}
\def\ei{\end{itemize}}
\def\bin{\begin{enumerate}}
\def\ein{\end{enumerate}}

\def\dc{\mathrm{I\!I\!I}}

\begin{document}
\title{Synthetic Random Flux Model in a periodically-driven optical lattice}
\author{Jan Major$^1$, Marcin P\l{}odzie\'n$^{1,2}$, Omjyoti Dutta$^3$ and Jakub Zakrzewski$^{1,4}$} 
\affiliation{\mbox{$^1$Instytut Fizyki imienia Mariana Smoluchowskiego, Uniwersytet Jagiello\'nski, \L{}ojasiewicza 11, 30-348 Krak\'ow, Poland}
\mbox{$^2$Eindhoven University of Technology, PO Box 513, 5600 MB Eindhoven, The Netherlands}
\mbox{$^3$Donostia International Physics Center (DIPC), Manuel de Lardizbal 4, E-20018 San Sebastian, Spain}
\mbox{$^4$Mark Kac Complex Systems Research Center, Jagiellonian University, \L{}ojasiewicza 11, 30-348  Krak\'ow, Poland}
}

\begin{abstract}
We propose a realization of a  synthetic Random Flux Model  in a two-dimensional optical lattice.
Starting from Bose-Hubbard Hamiltonian for two atom species we show how to use fast-periodic modulation of the system parameters to construct   
random gauge field. We investigate the transport properties of such a system and describe the impact of time-reversal symmetry breaking and correlations in disorder on Anderson localization length.
\end{abstract}
\maketitle
\section{Introduction}
The fractional quantum Hall effect (QHE) has been effectively described by the Chern-Simons field theory \cite{Kalmeyer1992,Halperin1993} in which quasiparticles are weakly interacting fermions constructed by attaching an even number of flux quanta to the electrons under a Chern-Simons transformation \cite{Jain1989}. In such a case fractional QHE is effectively mapped into integer QHE for the composite fermions suspended in an effective magnetic field. At filling factor $\nu_f=1/2$  the effective magnetic field vanishes and composite fermions are subject to random fluctuations of the gauge field induced by the ordinary impurities. In this context it is important to study the localization properties of non-interacting charged particles in the presence of a random magnetic field to understand the half-filling system. The problem of charged particles moving in a random magnetic field is also relevant to theoretical studies of high-$T_c$ models where gauge 
field fluctuations could significantly 
alter the critical temperature in high-$T_c$ superconductors \cite{Nagaosa1990}.

Anderson (strong) localization (AL) follows its precursor - a ``weak localization'' which describes a reduction of the conductivity due to constructive interferences between electronic paths and their time-reversed counterparts that hold at finite temperatures and a regime of small disorder. 
Due to the electron-electron and electron-phonon interactions a direct observation of Anderson localization in solid state systems is an impossible task and one has to rely on conductance measurements (for review see \cite{AndersonTransitionsReview}). Still  it has been directly observed in experiments with light \cite{Wiersma1997,Schwartz19999,Aegerter2007,Segev2013}, microwaves \cite{Pradhan2000}, ultrasound \cite{Hefei2008} and  ultracold quantum gases experiments \cite{Billy2008,Roati2008,Kondov2011,Gadway2011,Jendrzejewski2012,Semeghini2015}.
Scaling theory of localization predicts that in 2D noninteracting particles are Anderson localized, as an effect of quantum interference between time-reversal symmetric paths. Standard disordered systems with time-reversal symmetry has the coherent backward scattering which results in the weak localization.
The symmetry class of the problem could be changed (thus qualitatively changing results) for example
by addition of the spin-orbit coupling, which creates anti-localization correction leading to  an appearance of
the mobility edge in two dimensional systems \cite{Evangelou1987,Wang2014}. Another possible route is an addition of
the magnetic field which breaks the time-reversal symmetry, destroys the interference effects and results in the
suppression of the weak-localization correction \cite{Altshuler1980,Bergmann1984,Dupuis1992} leading to an increase of  the localization length. 
The case of Random Flux Model (RFM), where disorder appears as a random gauge field, is a subclass of systems with broken time-reversal symmetry. The existence of the mobility edge for RFM in two dimensions was for a long time a controversial issue with different predictions: some of them conclude that there exist extended states \cite{Lee1981,Sugiyama1993,Avishai1993,Liu1995,Kawarabayashi1995,Yakubo1996} while other conclude that the localization length in the vicinity of the band center is just extremely big so it could not be determined numerically \cite{Sheng1995,Xie1998,Furusaki1999}. 
RFM model with the diagonal disorder presents the interesting interplay between the two effects: as upon the appearance of random fluxes Anderson localization is weakened by  breaking of the time-reversal symmetry and simultaneously strengthened by the appearance of a flux disorder \cite{Chan1996,Verges1998}. 

 Cold atoms provide a particularly good environment for investigating AL. The ultracold atomic gases especially  'artificial crystals' -- the optical lattices, provide an unprecedented tunability of almost all parameters. The factors important for the localization such as the dimensionality of the system or the disorder distribution could be controlled. The interactions could be switched off with the help of the Feshbach resonances \cite{Chin2010}. The off-diagonal disorder and particularly random complex tunnelings -- equivalent to random fluxes of gauge field --  could be created by the means of the fast periodic modulation \cite{Kosior2015,Dutta2017}.

In this paper we propose an experimental scheme allowing the construction of the two-dimensional lattice system with synthetic random magnetic fields. We show that time-reversal symmetry breaking not always leads to anincrease of localization length.
We investigate  transport properties of systems, we propose, analyze a competition between the strengthening and the weakening of the localization by an introduction of random fluxes and present a simple toy model explaining an unexpectedly strong localization in some of cases with correlations.

The article is structured as follows: In  section \ref{sec:model} we describe the model we use: a two dimensional Bose-Hubbard Hamiltonian for two atomic species. The first specy, forming a diagonal disorder, is composed of  immobilized atoms randomly distributed in the lattice. A second one is formed by  mobile atoms that interact only with immobile atoms. Artificial gauge field is created by simultaneous fast periodic modulation of the mobile-immobile atoms interactions and a lattice height. Further in section \ref{sec:results} we present results of the numerical calculation of the localization length. We identify observed phenomena and present a simple model of transport through one plaquette to justify appearing discrepancies of localization length from the expected behavior. Finally in section \ref{sec:conclusions} we conclude.
\section{The model \label{sec:model}}
In order to create a disordered potential in the optical lattice we  consider randomly distributed \emph{frozen} particles
($\f$ superscript) with repulsive interactions.
Second species of atoms (\emph{mobile}), are non-interacting bosons experiencing \frozen{} atoms as a disorder potential.
For a deep lattice, the  system may be described by Bose-Hubbard model:
\begin{align}
H_0 &= \sum_{d}\sum_{\langle ij\rangle_d}(t a_i^\dagger a_j +t^\f a^{\f\dagger}_i a^\f_j)\nonumber\\ &+\sum_i\frac{U}{2}n_i (n_i-1)+\frac{U^\f}{2}n^\f_i (n^\f_i-1)+V n^\f_i n_i,
\end{align}
where $i$ is the lattice site, $a_i$ and $a^\dagger_i$ are bosonic annihilation and creation operators respectively, $n_i$ is a particle number operator, $t$ is a hopping amplitude between nearest neighbors, $U$ and $V$ are intraspecies and interspecies contact interaction strengths, respectively and $\langle ij\rangle_{d}$ denotes summing over the nearest neighbors in a direction $d$ which could be either $x$ or $y$. To obtain an appropriate  disordered potential we envision the following scenario: at the beginning only the \emph{frozen} particles are present in the lattice. By setting $t^\f \gtrsim U^\f$ we put the system into a deep superfluid state. Now,  the tunnelings are changed rapidly, for example by a fast increase of the lattice depth, the occupation of the lattice sites after the quench will be random and given by the Poisson distribution with the mean $\rho^\f$ -- mean occupation of the \frozen{} particles. Into such a prepared system the \emph{mobile} particles could be injected. As we assume 
that $t^\f=0$ and additionally that the \emph{mobile} particles interact only with the \frozen{} ones ($V\neq0$ and $U=0$, where the latter is obtained by the means of an optical or microwave Feshbach resonance \cite{Chin2010}) we get the Hamiltonian:
\begin{align}\label{eq:H_DD}
H_\mathrm{d} = t \sum_d\sum_{\langle ij\rangle_d}a_i^\dagger a_j + (V n^\f_i) n_i.
\end{align}
As a distribution of the \frozen{} particles is now fixed, we could treat $n^\f_i$ as a number and consequently the last term of the Hamiltonian \eqref{eq:H_DD} as just the on-site energy ($\epsilon_i = V n^\f_i$). This means that the Hamiltonian \eqref{eq:H_DD} describes a two dimensional system with the diagonal (on-site) disorder taken from the discrete Poisson distribution. In the next step we want to add a gauge field to this picture. As having the gauge field in the lattice is equivalent to adding complex phases to the tunnelings we will proceed with creating  complex phases using a fast periodic modulation of the lattice parameters. In our case we use the simultaneous modulation of the interspecies interaction $V\rightarrow V_0+V_1\sin(\omega t)$ and the tunneling rates $t\rightarrow t_0+t_1^\mathrm{(d)}f_\omega(t)$, where $\omega$ is the frequency of modulation and $f_\omega(t)$ is some periodic function. An important point is that we allow different modulations of the tunneling rates in different 
lattice directions. In an experiment the modulation of the interactions could be obtained by changing a magnetic field in the vicinity of the Feshbach resonance \cite{Rapp2012,Meinert2016}, while the tunneling rates could be changed by the modulation of the lattice depth.
The time dependent Hamiltonian reads:
\begin{align}
 H(t)&=\sum_d(t_0+t_1^\mathrm{(d)}f_\omega(t)) \sum_{\langle ij\rangle_d}a_i^\dagger a_j \nonumber\\&+ \sum_i((V_0+V_1\sin\omega t) n^\f_i) n_i.
\end{align}
$H(t)$ is time periodic so we  use Floquet theory \cite{Floquet1883,Bukov2015,Eckardt2015} to decouple fast \emph{micromotion} from long term dynamics described by a time independent effective Hamiltonian. Obtaining the exact effective Hamiltonian is usually a formidable task, but the approximate result could be calculated using the Magnus expansion \cite{Bukov2015,Kuwahara2016}, providing a series in powers of  $1/\omega$. In most of the cases the convergence rate of the series could be enhanced by a transformation to a rotating frame, alas in our case it is impossible as time dependent terms do not commute with each other. Nevertheless, we could make a partial transformation: 
\begin{align}
 \mathcal{U}=\exp\left(iV_1\cos\omega t /\omega \sum_i n_i^\f n_i\right),
\end{align}
which removes the time dependence from the on-site part of the Hamiltonian and more importantly takes system to a frame in which the modulation is a symmetric function of time (which makes the odd elements of the Magnus expansion identically $0$ \cite{Blanes2009}):
\begin{align}
 H'(t)&=\mathcal{U}H(t)\mathcal{U}^\dagger=\sum_i(V_0 n^\f_i) n_i\nonumber\\ &+\sum_d(t_0+t_1^\mathrm{(d)}f_\omega(t))\sum_{\langle ij\rangle_d}e^{i\frac{V_1}{\omega}(n^\f_j-n^\f_i)\cos\omega t}a_i^\dagger a_j.
\end{align}
In this frame the 0-th order of the Magnus expansion (simply a time average of the Hamiltonian) already gives the result with an error of the order of $1/\omega^2$\footnote{It is better than $O(1/\omega)$ which we will get if we have just proceeded with untransformed $H(t)$, however in the rotating frame we will have $O(1/\omega^4)$.}:
\begin{align}\label{eq:Heff}
H_\mathrm{eff}=\langle H'(t)\rangle_T =\sum_d\sum_{\langle ij\rangle_d}J^d_{ij}[f_\omega] a_i^\dagger a_j+(V_0 n^\f_i) n_i,
\end{align}
where $\langle.\rangle_T$ stands for a time averaging over period $T=2\pi/\omega$. The exact form of the effective tunneling rate $J_{ij}^d$ depends on the procedure of modulation $f_\omega$. We consider two different cases. First is the harmonic modulation:
\begin{align}\label{eq:HM}
J_{ij}^d[\cos\omega t] &= \langle (t_0+t_1^\mathrm{(d)}\cos\omega t)e^{i\frac{V_1}{\omega}(n^\f_j-n^\f_i)\cos\omega t} \rangle_T\nonumber\\ &= t_0\mathcal{J}_0\left(\frac{V_1}{\omega}(n^\f_j-n^\f_i)\right)+i t_1^{(d)}\mathcal{J}_1\left(\frac{V_1}{\omega}(n^\f_j-n^\f_i)\right),
\end{align}
where $\mathcal{J}_n(x)$ is $n$-th order Bessel function. If we set $t_1^{(d)}=\pm\sqrt{2}t_0$, eq.~\eqref{eq:HM} could be approximated as:
\begin{align}\label{eq:HMA}
 J_{ij}^d[\cos\omega t]&\approx \tilde{J_{ij}^d}[\cos\omega t]\nonumber\\&= t_0\exp\left(\pm i\tan^{-1}\left(\frac{V_1}{\omega}\left(n^\f_j-n^\f_i\right)\right)\right).
\end{align}

Although that approximation works only in the close vicinity of zero, especially for the phase, we will use it in calculations alongside the exact form \eqref{eq:HM}. $\tilde{J}_{ij}^d[cos\omega t]$ has several favorable features: its amplitude is always one, so only random fluxes are present (no random tunneling amplitudes); its phase depends nonlinearly on the argument, so in the case of the symmetric modulation in both directions ($t_1^{(x)}=t_1^{(y)}$) we could expect non-vanishing fluxes;  $\tan^{-1}$ saturates on $\pm\pi/2$ and consequently the fluxes takes values smaller than $2\pi$ ($2\pi$ is reached asymptotically for a very strong modulation) so the flux amplitude is monotonic function of the modulation strength.

A second option are periodic delta kicks $\dc_\omega(t)=\sum_n\delta(t+\frac{2\pi}{\omega}n)$, which gives:
\begin{align}\label{eq:DM}
 J_{ij}^d[\dc_\omega] = t_0\exp\left(\pm\frac{V_1}{\omega}\left(n^\f_j-n^\f_i\right)\right).
\end{align}
It has the desired property of the constant amplitude, unluckily its phase is changing linearly, so only for $t_1^{(x)}\neq t_1^{(y)}$ we will get nontrivial fluxes. Furthermore, its phase will wind up and in result it is possible to get smaller fluxes for stronger modulation/larger variation of particle number. Although this modulation procedure could  seem to be experimentally demanding, it could be easily approximated by the sum of the harmonic modulations: $t^\mathrm{(d)}_1(\cos\omega t+\cos2\omega t+\cos3\omega t+\ldots)$, for $t^\mathrm{(d)}_1=\pm 2 t_0$. Contrary the to previous case, here we have a very fast convergence both for the amplitude and the phase.

In our model the diagonal disorder is obviously correlated with the off-diagonal one, as both are taken from the same distribution of the \frozen{} particles. It is worth checking what impact on the localization this correlation has. To that end we consider also a different model in which we pick the on-site energies and tunnelings independently. Such a model with uncorrelated disorders is also possible to be experimentally realized, it could be created for example by using two different types of \frozen{} atoms. Yet another variation we consider is a model with solely diagonal disorder placed in a staggered gauge field. A reason for introducing that model is to distinguish effects due to  breaking of the time reversal symmetry from those created by an appearance of a new type of disorder.

\section{Results \label{sec:results}}

 \begin{figure}[t]
 \begin{center}
  \includegraphics[width=8cm]{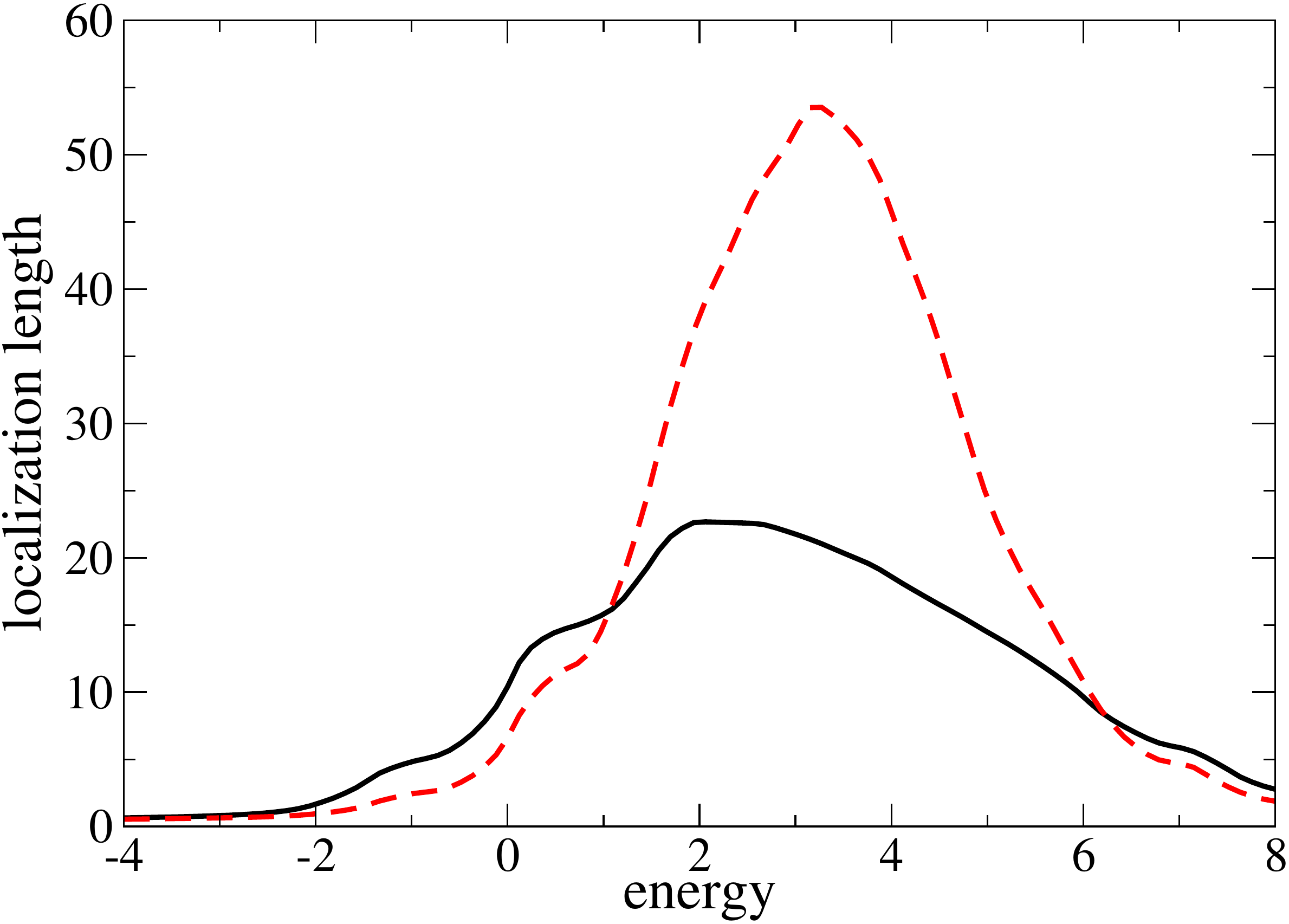}
  \caption{Anderson localization length (in units of the lattice constant) in function of the energy (in units of tunneling amplitude). Diagonal disorder is given by Poisson distribution of \frozen{} atoms (with mean $\rho_\f=2.5$) and interaction amplitude $V_0 = 1.5$. Black solid curve is for undriven system, red dashed is for delta modulation \eqref{eq:DM} with modulation parameter $V_1/\omega=1$ (for the correlated disorder case).}\label{Static_Disorder}  
 \end{center}
\end{figure}

\begin{figure}[ht]
\begin{center}
\includegraphics[width=8cm]{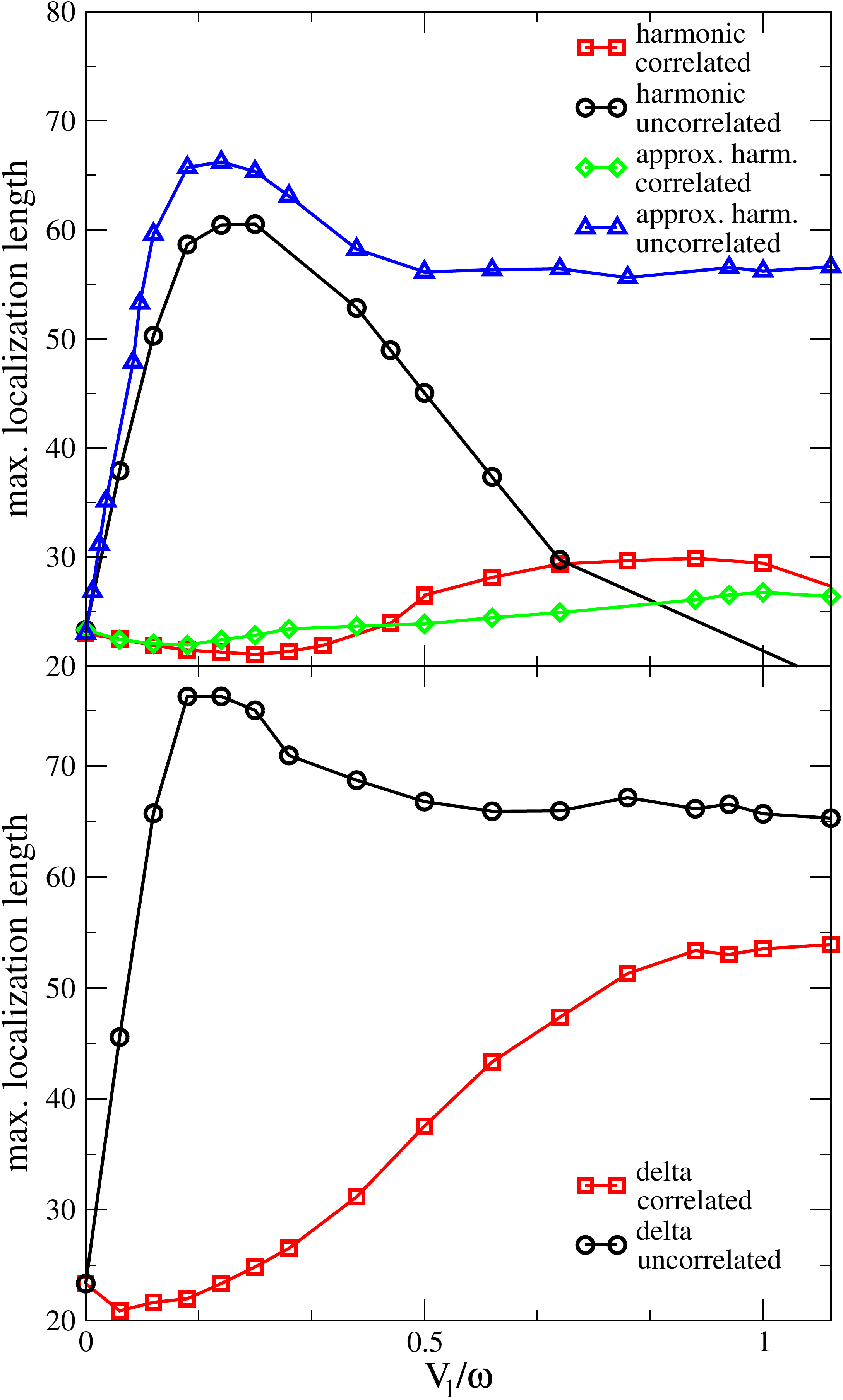}
\caption{The maximum localization length (in units of the lattice constant) as a function of the parameter of modulation $V_1/\omega$.
Top: Antisymmetric lattice modulation $t_1^\mathrm{(x)}=-t_1^\mathrm{(y)}$. Red squares and black discs shows results for harmonic modulation \eqref{eq:HM} for correlated and uncorrelated disorder respectively. Green diamonds and blue triangles are results for approximated harmonic modulation \eqref{eq:HMA} also for correlated and uncorrelated cases.
Bottom: Results for delta lattice modulation \eqref{eq:DM} Red squares are results for diagonal disorder correlated with off-diagonal, while black circles for uncorrelated case (lines are guides for the eye).  \label{fig:MainRes}
}
\end{center}
\end{figure}

In order to calculate Anderson localization length for system with diagonal and off-diagonal disorder we use the modified MacKinnon \& Kramer method \cite{MacKinnon1983,Kosior2015}. We numerically calculate two point Green's function in a quasi-1D stripe of size $M\times N$, where we increase $N$ to obtain a desired convergence. Each $i-th$ slice of a stripe is described by a one-dimensional Hamiltonian $H_i$ which is coupled to $i+1-th$ slice by $H_{i+1,j}$ matrix element.  Exponential decay of Green's function smallest eigenvalue allows us to extract the localization length, $\lambda_M(E),$ as a function of energy for a fixed disorder amplitude. Next, changing stripe width $M$ from 16 to 128 lattice sites we analyze the scaling behavior of $\lambda_M(E)/M$ and extract two dimensional Anderson localization length \cite{Abrahams1979}. 
 
In calculations we have used all three forms of effective tunneling described in preceding section. For harmonic modulation of lattice height $J_{ij}^d[\cos(\omega t)]$ \eqref{eq:HM} and its approximation $\tilde{J}_{ij}^d[\cos(\omega t)]$ \eqref{eq:HMA} we consider cases of symmetric $t_1^\mathrm{(x)}=t_1^\mathrm{(y)}$ as well as antisymmetric $t_1^\mathrm{(x)}=-t_1^\mathrm{(y)}$ modulation. For the delta modulation $J_{ij}^d[\dc_\omega]$ \eqref{eq:DM} only the antisymmetric case is calculated as the symmetric one gives no flux trivially. 
All presented results are calculated for interspecies interactions value $V_0=1.5$ (which effectively marks the scale of the on-site disorder). Qualitatively results for different $V_0$ values are similar but for smaller $V_0$ numerical errors grow due to a rapidly growing localization length, while for the stronger disorder the features become less distinctive. The mean density of  \emph{frozen} particles is fixed to $\rho_\f=2.5$. Due to a discrete character of the disorder used, there is a risk that for some energies and specific occupations of \frozen{} particles the resonant transport will occur and significantly alter the results. To check if it is an issue in our case we have done calculations for disorder taken from the folded normal distribution (with mean $\rho_\f$ and variance $\sqrt{\rho_\f}$), which greatly resembles the Poisson distribution. Results obtained in this way do not differ significantly.

Figure \ref{Static_Disorder} presents the localization length as a  function of energy for two cases: without the modulation and for a strong delta modulation. The behavior shown is typical for considered systems. We do not observe the mobility edge or separated extended states so we could rely on the maximal localization length (MLL) -- a maximal value of the localization length in the interval of energies studied -- as a good measure of the overall transport properties of the system for given parameters.

In Fig.~\ref{fig:MainRes} the MLL is plotted as a function of the modulation amplitude $V_1/\omega$. The upper panel shows results for 
the antisymmetric harmonic modulation (comparing the exact \eqref{eq:HM} and the approximate \eqref{eq:HMA} variants), in the lower panel results for delta modulation \eqref{eq:DM} are presented. In both cases models with correlated and uncorrelated disorder are considered.
Regardless the correlations between diagonal and off-diagonal disorder, the approximate results for the harmonic modulation \eqref{eq:HMA} agree  well with the exact results \eqref{eq:HM} for  $V_1/\omega$ up to $0.4$, see upper panel of   Fig.~\ref{fig:MainRes}. For modulations $V_1/\omega\gtrsim0.4$ the amplitude of expression \eqref{eq:HM} starts to significantly differ from $1$, and the disorder that appears in absolute values of the tunneling amplitudes seems to lower significantly the localization length.
\begin{figure}[t]
\begin{center}
\includegraphics[width=4cm]{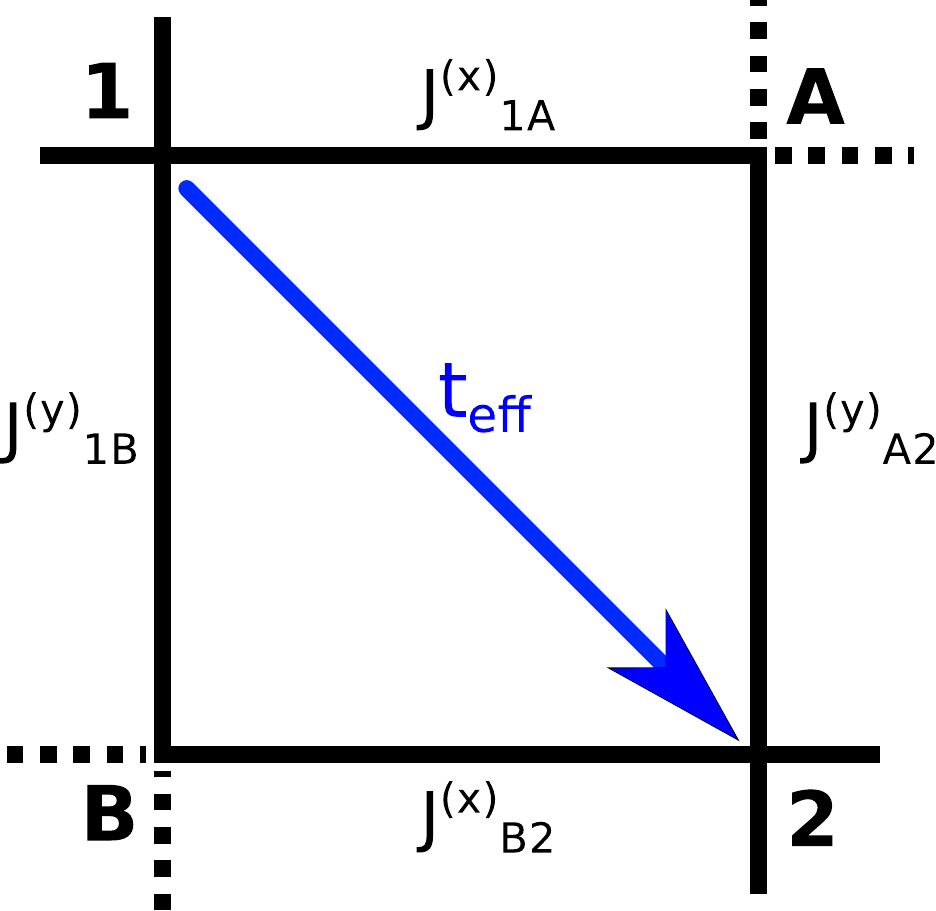}\\
\vspace{0.2cm}

\includegraphics[width=8cm]{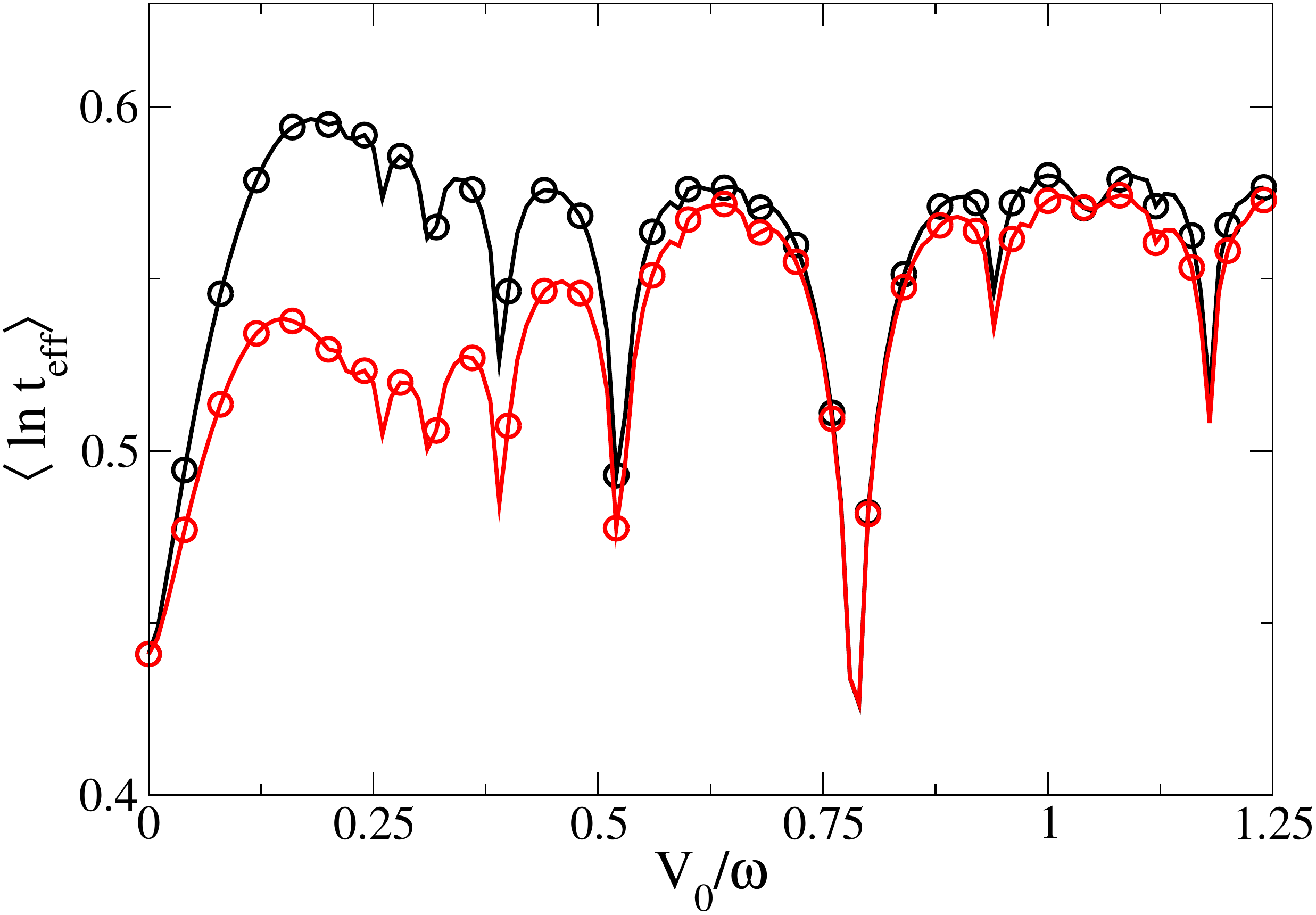}
\caption{
Top panel: A single lattice plaquette. Cutting the outer connections at sites A and B we could calculate tunneling through plaquette from site $1$ to site $2$ as given by \eqref{eq:teff}.
Bottom panel: The effective tunneling across the diagonal of the plaquette as a function of  $V_1/\omega$ for uncorrelated (black circles)  and correlated (red squares) of delta type lattice modulations at a single energy $E$ value. Similar behaior is observed at other energies.\label{fig:hist}}
\end{center}
\end{figure}
\begin{figure*}
 \begin{center}
  \includegraphics[width=16cm]{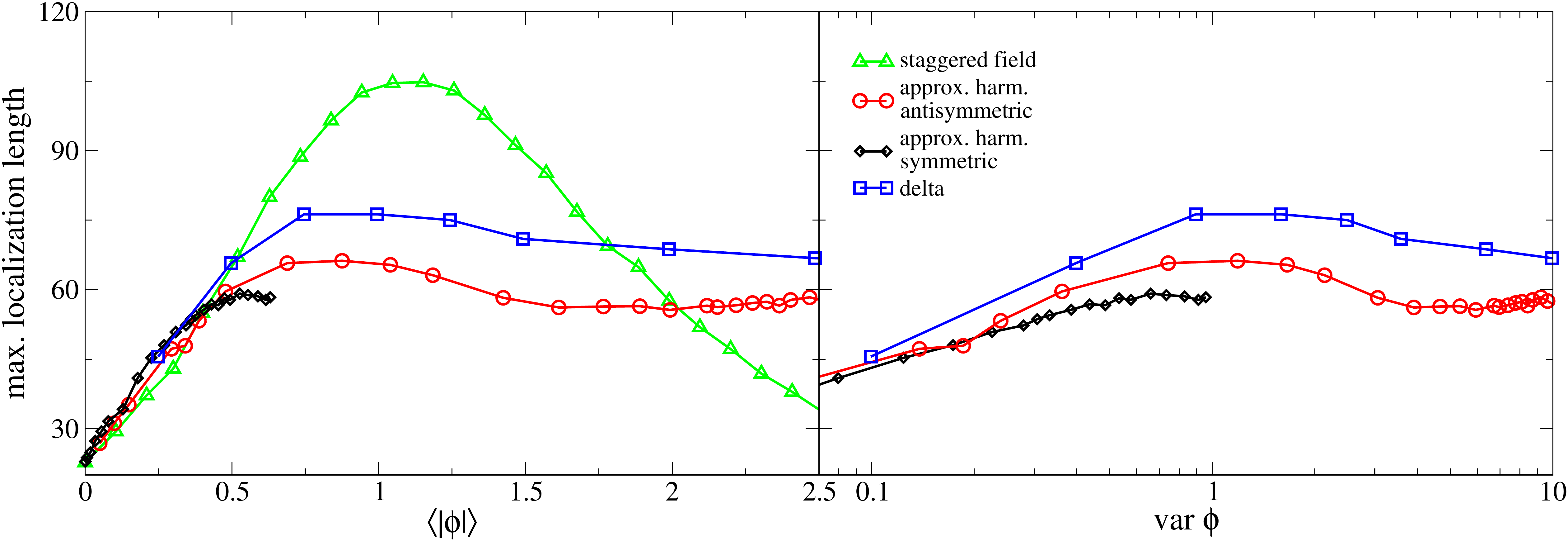}
 \end{center}
\caption{Maximum localization length as a function of the mean of the absolute flux through a single plaquette (left) 
 and as a function  of the variance of the flux through plaquette (right). Results are presented for the case of no correlation between  the diagonal and the off-diagonal disorder.   Black diamonds correspond to the approximate symmetric harmonic modulation \eqref{eq:HMA} while red circles to the approximate antisymmetric harmonic modulation ($t_1^\mathrm{(x)}=-t_1^\mathrm{(y)}$).   Blue squares stand for the delta modulation \eqref{eq:DM} and green triangles for the pure diagonal disorder in the staggered field (lines are drawn to  guide the eye).\label{fig:mean_var}
 }
\end{figure*}

The most striking of the observed effects is the large discrepancy between the results for the correlated and uncorrelated disorder, visible for all three considered effective tunnelings (both panels of Fig.~\ref{fig:MainRes}).  
In the uncorrelated case the MLL grows rapidly after the appearance of the random fluxes which is consistent with the growth of MLL expected when the time reversal symmetry is being broken.
Surprizingly, for the correlated case, MLL grows much slower or even shows a small decrease. 

In an attempt to understand this effect we analyse the transport through a single plaquette disconnected from the lattice (as in Fig.~\ref{fig:hist} top panel). The effective tunneling through such a structure is calculated to be: 
\begin{align}\label{eq:teff}
t_\mathrm{eff}=\frac{1}{E-V_0n^\f_A}J^{(y)}_{2A}J^{(x)}_{A1}+\frac{1}{E-V_0n^\f_B}J^{(x)}_{2B}J^{(y)}_{B1},
\end{align}
where the lattice sites are denoted as in the Fig.~\ref{fig:hist} and $E$ is energy of the state. We calculate $t_\mathrm{eff}$ for cases of the correlated and uncorrelated disorder (using the effective tunneling for the delta modulation case) and average it over disorder realizations. The results are shown in the lower panel of Fig.~\ref{fig:hist} versus the modulation parameter $V_1/\omega$.  As the states corresponding to MML have typically energies around $E=3.15$, this energy is chosen for calculation of $t_\mathrm{eff}$ (the results qualitatatively do not depend on $E$). We may observe qualitatatively similar behaviour of $t_{eff}$ for both correlated and uncorrelated disorder, for the latter the growth of $t_{eff}$ is significantly faster. For sufficiently large $V_1/\omega>0.7$ the difference between the two cases disappears.

Those results indicate  that understanding of the surprizing behavior observed for the correlated disorder cannot be obtained in the single plaquette model. Apparently the interference of different paths involving several plaquettes is responsible for the observed behavior for small and moderate $V_1/\omega$.
  
As the correlations between the diagonal disorder and the flux disorder affect localization properties strongly we should use results for the uncorrelated case to check the possible dependence of the MLL on the flux through a single plaquette.
Plotting MLL for the uncorrelated case as a function of a mean absolute flux through lattice plaquette  (Fig.~\ref{fig:mean_var} left panel), we can observe a similar behavior for the approximate harmonic modulation (for symmetric and antisymmetric versions) as well as for the delta modulation (we do not consider here the exact harmonic modulation as it gives also the disorder in absolute values of tunnelings which obscures the effects discussed). For smaller $V_1/\omega$ values those curves coincide with (also plotted) results for the diagonal disorder in the staggered field. This suggests that breaking of the time-reversal symmetry is a more important effect in this regime. For bigger fluxes the results for the disordered systems start to diverge from one for the staggered field -- this marks the region in which the random character of the fluxes gives a significant contribution to MLL. As we could see  in the right panel of Fig.~\ref{fig:mean_var} all three 
models  scale in similar manner also as a function of the variance of the flux.

For correlated disorder, the Anderson localization length can diverge for specific momenta. In the vicinity of such singularities there are intervals of momenta values in which the localization length is typically very large, allowing atoms with those momenta to leave the finite system.  In that way  the band-pass filter for momenta is formed as wavefunctions for  momenta outside of those, typically tiny, intervals, remain Anderson localized.
Such a mechanism has been proposed for BEC in speckle potential in 1D \cite{Plodzien2011} and in periodically-driven 1D optical lattice \cite{Kosior2015,Major2016}. 

The model presented in this paper can be utilized to construct a band-pass filter for the center of energy band in 2D.  Using the example of studied above static disorder with amplitude $V_0 = 1.5$ let us consider an optical lattice system of size $30\times30$. All atoms are Anderson localized and remain in the system. Now, applying harmonic antisymmetric modulation with $V_1/\omega = 1$ MLL for $E \approx 4$ increases to 55 lattice site (Fig.\ref{Static_Disorder}) and atoms with energy distribution centered at $E \approx 4$ can escape from the system.

\section{Conclusions\label{sec:conclusions}}
In this paper we have presented a method for creating  two-dimensional disordered system with artificial random gauge field for the ultracold atoms in the optical lattice using fast periodic modulations of atoms interaction. We showed that the time-reversal symmetry breaking does not necessary lead to increase of the Anderson localization length.
The presented model could be used to quantum simulation of high-$T_c$ superconductors models where scattering on random  gauge field could significantly lower the critical temperature \cite{Nagaosa1990}. 
\begin{acknowledgments} 
We are grateful to Dominique Delande for discussions.
This research has been supported by 
National Science Centre (Poland) under projects    2015/19/N/ST2/01677 (J.M.) and 2015/19/B/ST2/01028  (J.Z.).
J. Z. acknowledges also support by PL-Grid Infrastructure and EU  H2020-FETPROACT-2014 Project QUIC No.641122.
M. P. acknowledges support by the Foundation for Fundamental Research on Matter (FOM), and by the Netherlands Organisation for Scientific Research (NWO).
\end{acknowledgments}

\newpage
\begin{thebibliography}{45}
\expandafter\ifx\csname natexlab\endcsname\relax\def\natexlab#1{#1}\fi
\expandafter\ifx\csname bibnamefont\endcsname\relax
  \def\bibnamefont#1{#1}\fi
\expandafter\ifx\csname bibfnamefont\endcsname\relax
  \def\bibfnamefont#1{#1}\fi
\expandafter\ifx\csname citenamefont\endcsname\relax
  \def\citenamefont#1{#1}\fi
\expandafter\ifx\csname url\endcsname\relax
  \def\url#1{\texttt{#1}}\fi
\expandafter\ifx\csname urlprefix\endcsname\relax\def\urlprefix{URL }\fi
\providecommand{\bibinfo}[2]{#2}
\providecommand{\eprint}[2][]{\url{#2}}

\bibitem[{\citenamefont{Kalmeyer and Zhang}(1992)}]{Kalmeyer1992}
\bibinfo{author}{\bibfnamefont{V.}~\bibnamefont{Kalmeyer}} \bibnamefont{and}
  \bibinfo{author}{\bibfnamefont{S.-C.} \bibnamefont{Zhang}},
  \bibinfo{journal}{Phys. Rev. B} \textbf{\bibinfo{volume}{46}},
  \bibinfo{pages}{9889} (\bibinfo{year}{1992}).

\bibitem[{\citenamefont{Halperin et~al.}(1993)\citenamefont{Halperin, Lee, and
  Read}}]{Halperin1993}
\bibinfo{author}{\bibfnamefont{B.~I.} \bibnamefont{Halperin}},
  \bibinfo{author}{\bibfnamefont{P.~A.} \bibnamefont{Lee}}, \bibnamefont{and}
  \bibinfo{author}{\bibfnamefont{N.}~\bibnamefont{Read}},
  \bibinfo{journal}{Phys. Rev. B} \textbf{\bibinfo{volume}{47}},
  \bibinfo{pages}{7312} (\bibinfo{year}{1993}).

\bibitem[{\citenamefont{Jain}(1989)}]{Jain1989}
\bibinfo{author}{\bibfnamefont{J.~K.} \bibnamefont{Jain}},
  \bibinfo{journal}{Phys. Rev. Lett.} \textbf{\bibinfo{volume}{63}},
  \bibinfo{pages}{199} (\bibinfo{year}{1989}).

\bibitem[{\citenamefont{Nagaosa and Lee}(1990)}]{Nagaosa1990}
\bibinfo{author}{\bibfnamefont{N.}~\bibnamefont{Nagaosa}} \bibnamefont{and}
  \bibinfo{author}{\bibfnamefont{P.~A.} \bibnamefont{Lee}},
  \bibinfo{journal}{Phys. Rev. Lett.} \textbf{\bibinfo{volume}{64}},
  \bibinfo{pages}{2450} (\bibinfo{year}{1990}).

\bibitem[{\citenamefont{Evers and Mirlin}(2008)}]{AndersonTransitionsReview}
\bibinfo{author}{\bibfnamefont{F.}~\bibnamefont{Evers}} \bibnamefont{and}
  \bibinfo{author}{\bibfnamefont{A.~D.} \bibnamefont{Mirlin}},
  \bibinfo{journal}{Rev. Mod. Phys.} \textbf{\bibinfo{volume}{80}},
  \bibinfo{pages}{1355} (\bibinfo{year}{2008}).

\bibitem[{\citenamefont{Wiersma et~al.}(1997)\citenamefont{Wiersma, Bartolini,
  and Lagendijk}}]{Wiersma1997}
\bibinfo{author}{\bibfnamefont{D.}~\bibnamefont{Wiersma}},
  \bibinfo{author}{\bibfnamefont{P.}~\bibnamefont{Bartolini}},
  \bibnamefont{and} \bibinfo{author}{\bibfnamefont{A.~. R.~R.}
  \bibnamefont{Lagendijk}}, \bibinfo{journal}{Nature}
  \textbf{\bibinfo{volume}{390}}, \bibinfo{pages}{671} (\bibinfo{year}{1997}).

\bibitem[{\citenamefont{Schwartz et~al.}(1999)\citenamefont{Schwartz, Bartal,
  Fishman, and Segev}}]{Schwartz19999}
\bibinfo{author}{\bibfnamefont{T.}~\bibnamefont{Schwartz}},
  \bibinfo{author}{\bibfnamefont{G.}~\bibnamefont{Bartal}},
  \bibinfo{author}{\bibfnamefont{S.}~\bibnamefont{Fishman}}, \bibnamefont{and}
  \bibinfo{author}{\bibfnamefont{M.}~\bibnamefont{Segev}},
  \bibinfo{journal}{Nature} \textbf{\bibinfo{volume}{398}},
  \bibinfo{pages}{206} (\bibinfo{year}{1999}).

\bibitem[{\citenamefont{Aegerter et~al.}(2007)\citenamefont{Aegerter, Störzer,
  Fiebig, Bührer, and Maret}}]{Aegerter2007}
\bibinfo{author}{\bibfnamefont{C.~M.} \bibnamefont{Aegerter}},
  \bibinfo{author}{\bibfnamefont{M.}~\bibnamefont{Störzer}},
  \bibinfo{author}{\bibfnamefont{S.}~\bibnamefont{Fiebig}},
  \bibinfo{author}{\bibfnamefont{W.}~\bibnamefont{Bührer}}, \bibnamefont{and}
  \bibinfo{author}{\bibfnamefont{G.}~\bibnamefont{Maret}},
  \bibinfo{journal}{Nature} \textbf{\bibinfo{volume}{446}}, \bibinfo{pages}{52}
  (\bibinfo{year}{2007}).

\bibitem[{\citenamefont{Segev et~al.}(2013)\citenamefont{Segev, Silberger, and
  Christodoulides}}]{Segev2013}
\bibinfo{author}{\bibfnamefont{M.}~\bibnamefont{Segev}},
  \bibinfo{author}{\bibfnamefont{Y.}~\bibnamefont{Silberger}},
  \bibnamefont{and} \bibinfo{author}{\bibfnamefont{D.~N.}
  \bibnamefont{Christodoulides}}, \bibinfo{journal}{Nat. Photon.}
  \textbf{\bibinfo{volume}{7}}, \bibinfo{pages}{197} (\bibinfo{year}{2013}).

\bibitem[{\citenamefont{Pradhan and Sridar}(2000)}]{Pradhan2000}
\bibinfo{author}{\bibfnamefont{P.}~\bibnamefont{Pradhan}} \bibnamefont{and}
  \bibinfo{author}{\bibfnamefont{S.}~\bibnamefont{Sridar}},
  \bibinfo{journal}{Phys. Rev. Lett.} \textbf{\bibinfo{volume}{85}}, \bibinfo{pages}{2360}
  (\bibinfo{year}{2000}).

\bibitem[{\citenamefont{Hefei et~al.}(2008)\citenamefont{Hefei, Strybulevych,
  Page, Skipetrov, and Tiggelen}}]{Hefei2008}
\bibinfo{author}{\bibfnamefont{H.}~\bibnamefont{Hefei}},
  \bibinfo{author}{\bibfnamefont{A.}~\bibnamefont{Strybulevych}},
  \bibinfo{author}{\bibfnamefont{J.~H.} \bibnamefont{Page}},
  \bibinfo{author}{\bibfnamefont{E.}~\bibnamefont{Skipetrov}},
  \bibnamefont{and} \bibinfo{author}{\bibfnamefont{v.~B.}
  \bibnamefont{Tiggelen}}, \bibinfo{journal}{Nature Phys.}
  \textbf{\bibinfo{volume}{4}}, \bibinfo{pages}{945} (\bibinfo{year}{2008}).

\bibitem[{\citenamefont{Billy et~al.}(2008)\citenamefont{Billy, Josse, Zuol,
  Bernard, Hambrecth, Lugan, Clement, Sanchez-Palecia, Bouyer, and
  Aspect}}]{Billy2008}
\bibinfo{author}{\bibfnamefont{J.}~\bibnamefont{Billy}},
  \bibinfo{author}{\bibfnamefont{V.}~\bibnamefont{Josse}},
  \bibinfo{author}{\bibfnamefont{Z.}~\bibnamefont{Zuol}},
  \bibinfo{author}{\bibfnamefont{A.}~\bibnamefont{Bernard}},
  \bibinfo{author}{\bibfnamefont{B.}~\bibnamefont{Hambrecth}},
  \bibinfo{author}{\bibfnamefont{P.}~\bibnamefont{Lugan}},
  \bibinfo{author}{\bibfnamefont{D.}~\bibnamefont{Clement}},
  \bibinfo{author}{\bibfnamefont{L.}~\bibnamefont{Sanchez-Palecia}},
  \bibinfo{author}{\bibfnamefont{P.}~\bibnamefont{Bouyer}}, \bibnamefont{and}
  \bibinfo{author}{\bibfnamefont{A.}~\bibnamefont{Aspect}},
  \bibinfo{journal}{Nature} \textbf{\bibinfo{volume}{453}},
  \bibinfo{pages}{891} (\bibinfo{year}{2008}).

\bibitem[{\citenamefont{Roati et~al.}(2008)\citenamefont{Roati, D'Errico,
  Fallani, Fattori, Fort, Zaccanti, Modugno, Modugno, and Massimo}}]{Roati2008}
\bibinfo{author}{\bibfnamefont{G.}~\bibnamefont{Roati}},
  \bibinfo{author}{\bibfnamefont{C.}~\bibnamefont{D'Errico}},
  \bibinfo{author}{\bibfnamefont{L.}~\bibnamefont{Fallani}},
  \bibinfo{author}{\bibfnamefont{M.}~\bibnamefont{Fattori}},
  \bibinfo{author}{\bibfnamefont{C.}~\bibnamefont{Fort}},
  \bibinfo{author}{\bibfnamefont{M.}~\bibnamefont{Zaccanti}},
  \bibinfo{author}{\bibfnamefont{G.}~\bibnamefont{Modugno}},
  \bibinfo{author}{\bibfnamefont{M.}~\bibnamefont{Modugno}}, \bibnamefont{and}
  \bibinfo{author}{\bibfnamefont{I.}~\bibnamefont{Massimo}},
  \bibinfo{journal}{Nature} \textbf{\bibinfo{volume}{453}},
  \bibinfo{pages}{895} (\bibinfo{year}{2008}).

\bibitem[{\citenamefont{Kondov et~al.}(2011)\citenamefont{Kondov, McGehee,
  Zirbel, and DeMarco}}]{Kondov2011}
\bibinfo{author}{\bibfnamefont{S.~S.} \bibnamefont{Kondov}},
  \bibinfo{author}{\bibfnamefont{W.~R.} \bibnamefont{McGehee}},
  \bibinfo{author}{\bibfnamefont{J.~J.} \bibnamefont{Zirbel}},
  \bibnamefont{and} \bibinfo{author}{\bibfnamefont{B.}~\bibnamefont{DeMarco}},
  \bibinfo{journal}{Science} \textbf{\bibinfo{volume}{334}},
  \bibinfo{pages}{66} (\bibinfo{year}{2011}).

\bibitem[{\citenamefont{Gadway et~al.}(2011)\citenamefont{Gadway, Pertot,
  Reeves, Vogt, and Schneble}}]{Gadway2011}
\bibinfo{author}{\bibfnamefont{B.}~\bibnamefont{Gadway}},
  \bibinfo{author}{\bibfnamefont{D.}~\bibnamefont{Pertot}},
  \bibinfo{author}{\bibfnamefont{J.}~\bibnamefont{Reeves}},
  \bibinfo{author}{\bibfnamefont{M.}~\bibnamefont{Vogt}}, \bibnamefont{and}
  \bibinfo{author}{\bibfnamefont{D.}~\bibnamefont{Schneble}},
  \bibinfo{journal}{Phys. Rev. Lett.} \textbf{\bibinfo{volume}{107}},
  \bibinfo{pages}{145306} (\bibinfo{year}{2011}),
  \urlprefix\url{https://link.aps.org/doi/10.1103/PhysRevLett.107.145306}.

\bibitem[{\citenamefont{Jendrzejewski et~al.}(2012)\citenamefont{Jendrzejewski,
  Bernard, Müller, Cheinet, Josse, Piraud, Pezzé, Sanchez-Palencia, Aspect,
  and Bouyer}}]{Jendrzejewski2012}
\bibinfo{author}{\bibfnamefont{F.}~\bibnamefont{Jendrzejewski}},
  \bibinfo{author}{\bibfnamefont{A.}~\bibnamefont{Bernard}},
  \bibinfo{author}{\bibfnamefont{K.}~\bibnamefont{Müller}},
  \bibinfo{author}{\bibfnamefont{P.}~\bibnamefont{Cheinet}},
  \bibinfo{author}{\bibfnamefont{V.}~\bibnamefont{Josse}},
  \bibinfo{author}{\bibfnamefont{M.}~\bibnamefont{Piraud}},
  \bibinfo{author}{\bibfnamefont{L.}~\bibnamefont{Pezzé}},
  \bibinfo{author}{\bibfnamefont{L.}~\bibnamefont{Sanchez-Palencia}},
  \bibinfo{author}{\bibfnamefont{A.}~\bibnamefont{Aspect}}, \bibnamefont{and}
  \bibinfo{author}{\bibfnamefont{P.}~\bibnamefont{Bouyer}},
  \bibinfo{journal}{Nature Phys.} \textbf{\bibinfo{volume}{8}},
  \bibinfo{pages}{398} (\bibinfo{year}{2012}).

\bibitem[{\citenamefont{Semeghini et~al.}(2015)\citenamefont{Semeghini,
  Landini, Castilho, Roy, Spagnolli, Trenkwalder, Fattori, Inguscio, and
  Modugno}}]{Semeghini2015}
\bibinfo{author}{\bibfnamefont{G.}~\bibnamefont{Semeghini}},
  \bibinfo{author}{\bibfnamefont{M.}~\bibnamefont{Landini}},
  \bibinfo{author}{\bibfnamefont{P.}~\bibnamefont{Castilho}},
  \bibinfo{author}{\bibfnamefont{S.}~\bibnamefont{Roy}},
  \bibinfo{author}{\bibfnamefont{G.}~\bibnamefont{Spagnolli}},
  \bibinfo{author}{\bibfnamefont{A.}~\bibnamefont{Trenkwalder}},
  \bibinfo{author}{\bibfnamefont{M.}~\bibnamefont{Fattori}},
  \bibinfo{author}{\bibfnamefont{M.}~\bibnamefont{Inguscio}}, \bibnamefont{and}
  \bibinfo{author}{\bibfnamefont{G.}~\bibnamefont{Modugno}},
  \bibinfo{journal}{Nature Physics} \textbf{\bibinfo{volume}{11}},
  \bibinfo{pages}{554} (\bibinfo{year}{2015}).

\bibitem[{\citenamefont{Evangelou and Ziman}(1987)\citenamefont{Evangelou, Ziman}}]{Evangelou1987}
\bibinfo{author}{\bibfnamefont{S.~N.}~\bibnamefont{Evangelou}},
  \bibinfo{author}{\bibfnamefont{T.}~\bibnamefont{Ziman}},
  \bibinfo{journal}{Journal of Physics C: Solid State Physics} \textbf{\bibinfo{volume}{20}},
  \bibinfo{pages}{L235} (\bibinfo{year}{1987}).
  
\bibitem[{\citenamefont{Wang et~al.}(2014)\citenamefont{Wang, Su, Avishai, Meir, Wang}}]{Wang2014}
\bibinfo{author}{\bibfnamefont{C.}~\bibnamefont{Wang}} \bibnamefont{and}
  \bibinfo{author}{\bibfnamefont{Y.}~\bibnamefont{Su}}\bibnamefont{and}
  \bibinfo{author}{\bibfnamefont{Y.}~\bibnamefont{Avishai}}\bibnamefont{and}
  \bibinfo{author}{\bibfnamefont{Y.}~\bibnamefont{Meir}}\bibnamefont{and}
  \bibinfo{author}{\bibfnamefont{X.~R.}~\bibnamefont{Wang}},
  \bibinfo{journal}{ArXiv e-prints},
  \bibinfo{pages}{1411.4838} (\bibinfo{year}{2014}).
  
  
\bibitem[{\citenamefont{Altshuler et~al.}(1980)\citenamefont{Altshuler,
  Khmel'nitzkii, Larkin, and Lee}}]{Altshuler1980}
\bibinfo{author}{\bibfnamefont{B.~L.} \bibnamefont{Altshuler}},
  \bibinfo{author}{\bibfnamefont{D.}~\bibnamefont{Khmel'nitzkii}},
  \bibinfo{author}{\bibfnamefont{A.~I.} \bibnamefont{Larkin}},
  \bibnamefont{and} \bibinfo{author}{\bibfnamefont{P.~A.} \bibnamefont{Lee}},
  \bibinfo{journal}{Phys. Rev. B} \textbf{\bibinfo{volume}{22}},
  \bibinfo{pages}{5142} (\bibinfo{year}{1980}).

\bibitem[{\citenamefont{Bergmann}(1984)}]{Bergmann1984}
\bibinfo{author}{\bibfnamefont{G.}~\bibnamefont{Bergmann}},
  \bibinfo{journal}{Phys. Rep.} \textbf{\bibinfo{volume}{107}},
  \bibinfo{pages}{1 } (\bibinfo{year}{1984}).

\bibitem[{\citenamefont{Dupuis and Montambaux}(1992)}]{Dupuis1992}
\bibinfo{author}{\bibfnamefont{N.}~\bibnamefont{Dupuis}} \bibnamefont{and}
  \bibinfo{author}{\bibfnamefont{G.}~\bibnamefont{Montambaux}},
  \bibinfo{journal}{Phys. Rev. Lett.} \textbf{\bibinfo{volume}{68}},
  \bibinfo{pages}{357} (\bibinfo{year}{1992}).

\bibitem[{\citenamefont{Lee and Fisher}(1981)}]{Lee1981}
\bibinfo{author}{\bibfnamefont{P.~A.} \bibnamefont{Lee}} \bibnamefont{and}
  \bibinfo{author}{\bibfnamefont{D.~S.} \bibnamefont{Fisher}},
  \bibinfo{journal}{Phys. Rev. Lett.} \textbf{\bibinfo{volume}{47}},
  \bibinfo{pages}{882} (\bibinfo{year}{1981}).

\bibitem[{\citenamefont{Sugiyama and Nagaosa}(1993)}]{Sugiyama1993}
\bibinfo{author}{\bibfnamefont{T.}~\bibnamefont{Sugiyama}} \bibnamefont{and}
  \bibinfo{author}{\bibfnamefont{N.}~\bibnamefont{Nagaosa}},
  \bibinfo{journal}{Phys. Rev. Lett.} \textbf{\bibinfo{volume}{70}},
  \bibinfo{pages}{1980} (\bibinfo{year}{1993}).

\bibitem[{\citenamefont{Avishai et~al.}(1993)\citenamefont{Avishai, Hatsugai,
  and Kohmoto}}]{Avishai1993}
\bibinfo{author}{\bibfnamefont{Y.}~\bibnamefont{Avishai}},
  \bibinfo{author}{\bibfnamefont{Y.}~\bibnamefont{Hatsugai}}, \bibnamefont{and}
  \bibinfo{author}{\bibfnamefont{M.}~\bibnamefont{Kohmoto}},
  \bibinfo{journal}{Phys. Rev. B} \textbf{\bibinfo{volume}{47}},
  \bibinfo{pages}{9561} (\bibinfo{year}{1993}).

\bibitem[{\citenamefont{Liu et~al.}(1995)\citenamefont{Liu, Xie, Das~Sarma, and
  Zhang}}]{Liu1995}
\bibinfo{author}{\bibfnamefont{D.~Z.} \bibnamefont{Liu}},
  \bibinfo{author}{\bibfnamefont{X.~C.} \bibnamefont{Xie}},
  \bibinfo{author}{\bibfnamefont{S.}~\bibnamefont{Das~Sarma}},
  \bibnamefont{and} \bibinfo{author}{\bibfnamefont{S.~C.} \bibnamefont{Zhang}},
  \bibinfo{journal}{Phys. Rev. B} \textbf{\bibinfo{volume}{52}},
  \bibinfo{pages}{5858} (\bibinfo{year}{1995}).

\bibitem[{\citenamefont{Kawarabayashi and Ohtsuki}(1995)}]{Kawarabayashi1995}
\bibinfo{author}{\bibfnamefont{T.}~\bibnamefont{Kawarabayashi}}
  \bibnamefont{and} \bibinfo{author}{\bibfnamefont{T.}~\bibnamefont{Ohtsuki}},
  \bibinfo{journal}{Phys. Rev. B} \textbf{\bibinfo{volume}{51}},
  \bibinfo{pages}{10897} (\bibinfo{year}{1995}).

\bibitem[{\citenamefont{Yakubo and Goto}(1996)}]{Yakubo1996}
\bibinfo{author}{\bibfnamefont{K.}~\bibnamefont{Yakubo}} \bibnamefont{and}
  \bibinfo{author}{\bibfnamefont{Y.}~\bibnamefont{Goto}},
  \bibinfo{journal}{Phys. Rev. B} \textbf{\bibinfo{volume}{54}},
  \bibinfo{pages}{13432} (\bibinfo{year}{1996}).

\bibitem[{\citenamefont{Sheng and Weng}(1995)}]{Sheng1995}
\bibinfo{author}{\bibfnamefont{D.~N.} \bibnamefont{Sheng}} \bibnamefont{and}
  \bibinfo{author}{\bibfnamefont{Z.~Y.} \bibnamefont{Weng}},
  \bibinfo{journal}{Phys. Rev. Lett.} \textbf{\bibinfo{volume}{75}},
  \bibinfo{pages}{2388} (\bibinfo{year}{1995}).

\bibitem[{\citenamefont{Xie et~al.}(1998)\citenamefont{Xie, Wang, and
  Liu}}]{Xie1998}
\bibinfo{author}{\bibfnamefont{X.~C.} \bibnamefont{Xie}},
  \bibinfo{author}{\bibfnamefont{X.~R.} \bibnamefont{Wang}}, \bibnamefont{and}
  \bibinfo{author}{\bibfnamefont{D.~Z.} \bibnamefont{Liu}},
  \bibinfo{journal}{Phys. Rev. Lett.} \textbf{\bibinfo{volume}{80}},
  \bibinfo{pages}{3563} (\bibinfo{year}{1998}).

\bibitem[{\citenamefont{Furusaki}(1999)}]{Furusaki1999}
\bibinfo{author}{\bibfnamefont{A.}~\bibnamefont{Furusaki}},
  \bibinfo{journal}{Phys. Rev. Lett.} \textbf{\bibinfo{volume}{82}},
  \bibinfo{pages}{604} (\bibinfo{year}{1999}).

\bibitem[{\citenamefont{Chan et~al.}(1996)\citenamefont{Chan, Wang, and
  Xie}}]{Chan1996}
\bibinfo{author}{\bibfnamefont{W.~L.} \bibnamefont{Chan}},
  \bibinfo{author}{\bibfnamefont{X.~R.} \bibnamefont{Wang}}, \bibnamefont{and}
  \bibinfo{author}{\bibfnamefont{X.~C.} \bibnamefont{Xie}},
  \bibinfo{journal}{Phys. Rev. B} \textbf{\bibinfo{volume}{54}},
  \bibinfo{pages}{11213} (\bibinfo{year}{1996}).

\bibitem[{\citenamefont{Verg\'es}(1998)}]{Verges1998}
\bibinfo{author}{\bibfnamefont{J.~A.} \bibnamefont{Verg\'es}},
  \bibinfo{journal}{Phys. Rev. B} \textbf{\bibinfo{volume}{57}},
  \bibinfo{pages}{870} (\bibinfo{year}{1998}).

\bibitem[{\citenamefont{Chin et~al.}(2010)\citenamefont{Chin, Grimm, Julienne,
  and Tiesinga}}]{Chin2010}
\bibinfo{author}{\bibfnamefont{C.}~\bibnamefont{Chin}},
  \bibinfo{author}{\bibfnamefont{R.}~\bibnamefont{Grimm}},
  \bibinfo{author}{\bibfnamefont{P.}~\bibnamefont{Julienne}}, \bibnamefont{and}
  \bibinfo{author}{\bibfnamefont{E.}~\bibnamefont{Tiesinga}},
  \bibinfo{journal}{Rev. Mod. Phys.} \textbf{\bibinfo{volume}{82}},
  \bibinfo{pages}{1225} (\bibinfo{year}{2010}).

\bibitem[{\citenamefont{Rapp et~al.}(2012)\citenamefont{Rapp, Deng, and
  Santos}}]{Rapp2012}
\bibinfo{author}{\bibfnamefont{A.}~\bibnamefont{Rapp}},
  \bibinfo{author}{\bibfnamefont{X.}~\bibnamefont{Deng}}, \bibnamefont{and}
  \bibinfo{author}{\bibfnamefont{L.}~\bibnamefont{Santos}},
  \bibinfo{journal}{Phys. Rev. Lett.} \textbf{\bibinfo{volume}{109}},
  \bibinfo{pages}{203005} (\bibinfo{year}{2012}).

\bibitem[{\citenamefont{Meinert et~al.}(2016)\citenamefont{Meinert, Mark,
  Lauber, Daley, and N\"agerl}}]{Meinert2016}
\bibinfo{author}{\bibfnamefont{F.}~\bibnamefont{Meinert}},
  \bibinfo{author}{\bibfnamefont{M.~J.} \bibnamefont{Mark}},
  \bibinfo{author}{\bibfnamefont{K.}~\bibnamefont{Lauber}},
  \bibinfo{author}{\bibfnamefont{A.~J.} \bibnamefont{Daley}}, \bibnamefont{and}
  \bibinfo{author}{\bibfnamefont{H.-C.} \bibnamefont{N\"agerl}},
  \bibinfo{journal}{Phys. Rev. Lett.} \textbf{\bibinfo{volume}{116}},
  \bibinfo{pages}{205301} (\bibinfo{year}{2016}),
  \urlprefix\url{https://link.aps.org/doi/10.1103/PhysRevLett.116.205301}.

\bibitem[{\citenamefont{Floquet}(1883)}]{Floquet1883}
\bibinfo{author}{\bibfnamefont{G.}~\bibnamefont{Floquet}},
  \bibinfo{journal}{Ann. Sci. Ec. norm. Super.} \textbf{\bibinfo{volume}{12}},
  \bibinfo{pages}{47} (\bibinfo{year}{1883}).

\bibitem[{\citenamefont{Bukov et~al.}(2015)\citenamefont{Bukov, D'Alessio, and
  Polkovnikov}}]{Bukov2015}
\bibinfo{author}{\bibfnamefont{M.}~\bibnamefont{Bukov}},
  \bibinfo{author}{\bibfnamefont{L.}~\bibnamefont{D'Alessio}},
  \bibnamefont{and}
  \bibinfo{author}{\bibfnamefont{A.}~\bibnamefont{Polkovnikov}},
  \bibinfo{journal}{Adv. Phys.} \textbf{\bibinfo{volume}{64}},
  \bibinfo{pages}{139} (\bibinfo{year}{2015}).

\bibitem[{\citenamefont{Eckardt and Anisimovas}(2015)}]{Eckardt2015}
\bibinfo{author}{\bibfnamefont{A.}~\bibnamefont{Eckardt}} \bibnamefont{and}
  \bibinfo{author}{\bibfnamefont{E.}~\bibnamefont{Anisimovas}},
  \bibinfo{journal}{New J. Phys.} \textbf{\bibinfo{volume}{17}},
  \bibinfo{pages}{093039} (\bibinfo{year}{2015}).

\bibitem[{\citenamefont{Kuwahara et~al.}(2016)\citenamefont{Kuwahara, Mori, and
  Saito}}]{Kuwahara2016}
\bibinfo{author}{\bibfnamefont{T.}~\bibnamefont{Kuwahara}},
  \bibinfo{author}{\bibfnamefont{T.}~\bibnamefont{Mori}}, \bibnamefont{and}
  \bibinfo{author}{\bibfnamefont{K.}~\bibnamefont{Saito}},
  \bibinfo{journal}{Annals of Physics} \textbf{\bibinfo{volume}{367}},
  \bibinfo{pages}{96 } (\bibinfo{year}{2016}), ISSN \bibinfo{issn}{0003-4916},
  \urlprefix\url{http://www.sciencedirect.com/science/article/pii/S0003491616000142}.

\bibitem[{\citenamefont{Blanes et~al.}(2009)\citenamefont{Blanes, Casas, Oteo,
  and Ros}}]{Blanes2009}
\bibinfo{author}{\bibfnamefont{S.}~\bibnamefont{Blanes}},
  \bibinfo{author}{\bibfnamefont{F.}~\bibnamefont{Casas}},
  \bibinfo{author}{\bibfnamefont{J.}~\bibnamefont{Oteo}}, \bibnamefont{and}
  \bibinfo{author}{\bibfnamefont{J.}~\bibnamefont{Ros}},
  \bibinfo{journal}{Phys. Rep.} \textbf{\bibinfo{volume}{470}},
  \bibinfo{pages}{151 } (\bibinfo{year}{2009}).

\bibitem[{Note1()}]{Note1}
Note1, \bibinfo{note}{it is better than $O(1/\omega )$ which we will get if we
  have just proceeded with untransformed $H(t)$, however in the rotating frame
  we will have $O(1/\omega ^4)$.}

\bibitem[{\citenamefont{MacKinnon and Kramer}(1983)}]{MacKinnon1983}
\bibinfo{author}{\bibfnamefont{A.}~\bibnamefont{MacKinnon}} \bibnamefont{and}
  \bibinfo{author}{\bibfnamefont{B.}~\bibnamefont{Kramer}},
  \bibinfo{journal}{Z. Phys. B} \textbf{\bibinfo{volume}{53}},
  \bibinfo{pages}{1} (\bibinfo{year}{1983}).

\bibitem[{\citenamefont{Kosior et~al.}(2015)\citenamefont{Kosior, Major,
  P\l{}odzie\'{n}, and Zakrzewski}}]{Kosior2015}
\bibinfo{author}{\bibfnamefont{A.}~\bibnamefont{Kosior}},
  \bibinfo{author}{\bibfnamefont{J.}~\bibnamefont{Major}},
  \bibinfo{author}{\bibfnamefont{M.}~\bibnamefont{P\l{}odzie\'{n}}},
  \bibnamefont{and}
  \bibinfo{author}{\bibfnamefont{J.}~\bibnamefont{Zakrzewski}},
  \bibinfo{journal}{Phys. Rev. A} \textbf{\bibinfo{volume}{92}},
  \bibinfo{pages}{023606} (\bibinfo{year}{2015}).

  \bibitem[{\citenamefont{Dutta et~al.}(2017)\citenamefont{Dutta, Tagliacozzo, Lewenstein,
   and Zakrzewski}}]{Dutta2017}
\bibinfo{author}{\bibfnamefont{O.}~\bibnamefont{Dutta}},
  \bibinfo{author}{\bibfnamefont{L.}~\bibnamefont{Tagliacozzo}},
  \bibinfo{author}{\bibfnamefont{M.}~\bibnamefont{Lewenstein}},
  \bibnamefont{and}
  \bibinfo{author}{\bibfnamefont{J.}~\bibnamefont{Zakrzewski}},
  \bibinfo{journal}{Phys. Rev. A} \textbf{\bibinfo{volume}{95}},
  \bibinfo{pages}{053608} (\bibinfo{year}{2017}).

\bibitem[{\citenamefont{Abrahams et~al.}(1979)\citenamefont{Abrahams, Anderson,
  Licciardello, and Ramakrishnan}}]{Abrahams1979}
\bibinfo{author}{\bibfnamefont{E.}~\bibnamefont{Abrahams}},
  \bibinfo{author}{\bibfnamefont{P.~W.} \bibnamefont{Anderson}},
  \bibinfo{author}{\bibfnamefont{D.~C.} \bibnamefont{Licciardello}},
  \bibnamefont{and} \bibinfo{author}{\bibfnamefont{T.~V.}
  \bibnamefont{Ramakrishnan}}, \bibinfo{journal}{Phys. Rev. Lett.}
  \textbf{\bibinfo{volume}{42}}, \bibinfo{pages}{673} (\bibinfo{year}{1979}).

\bibitem[{\citenamefont{P\l{}odzie\ifmmode~\acute{n}\else \'{n}\fi{} and
  Sacha}(2011)}]{Plodzien2011}
\bibinfo{author}{\bibfnamefont{M.}~\bibnamefont{P\l{}odzie\ifmmode~\acute{n}\else
  \'{n}\fi{}}} \bibnamefont{and}
  \bibinfo{author}{\bibfnamefont{K.}~\bibnamefont{Sacha}},
  \bibinfo{journal}{Phys. Rev. A} \textbf{\bibinfo{volume}{84}},
  \bibinfo{pages}{023624} (\bibinfo{year}{2011}).

\bibitem[{\citenamefont{Major}(2016)}]{Major2016}
\bibinfo{author}{\bibfnamefont{J.}~\bibnamefont{Major}},
  \bibinfo{journal}{Phys. Rev. A} \textbf{\bibinfo{volume}{94}},
  \bibinfo{pages}{053613} (\bibinfo{year}{2016}).

\end{thebibliography}

\end{document}